\documentstyle[prd,aps,epsfig]{revtex}
\begin{document}
\title{Non-local anomaly of the axial-vector current for bound states}
\author{Dmitri Melikhov and Berthold Stech}
\address{Institut f\"ur Theoretische Physik, Universit\"at Heidelberg,  
Philosophenweg 16, D-69120, Heidelberg, Germany}
\maketitle
\begin{abstract}
We demonstrate that the amplitude 
$\langle\rho\gamma|\partial^\nu (\bar q\gamma_\nu \gamma_5 q)|0\rangle$ 
does not vanish in the limit of zero quark masses. 
This represents a new kind of violation of the classical equation of motion 
for the axial current and should be interpreted as the axial anomaly for 
bound states. The anomaly emerges in spite of the fact that the one loop 
integrals are ultraviolet-finite as guaranteed by the finite-size of  
bound-state wave functions. As a result, the amplitude behaves like 
$\sim 1/p^2$ in the limit of a large momentum $p$ of the current. 
This is to be compared with the amplitude 
$\langle \gamma\gamma|\partial^\nu (\bar q\gamma_\nu \gamma_5 q)|0\rangle$ 
which remains finite in the limit $p^2\to\infty$. 

The observed effect requires the modification of the classical equation of 
motion of the axial-vector current by non-local operators. The non-local 
axial anomaly is a general phenomenon which is effective for axial-vector 
currents interacting with spin-1 bound states. 

\vspace{.1cm}
\end{abstract}

\vspace{.4cm}
The analysis of two-photon decays of pseudoscalar mesons in the late 40-s \cite{st} led to 
the discovery of the famous axial anomaly \cite{abj}: the divergence of the axial vector 
current violates the classical equation of motion and does not vanish in the limit of zero 
fermion masses. 
For a quark of mass $m_q$ and charge $eQ_q$ the properly modified equation of motion 
contains a local anomalous term and has the form\footnote{
We use the following notations: 
$e=\sqrt{4\pi\alpha_{\rm em}}$, 
$\gamma^5=i\gamma^0\gamma^1\gamma^2\gamma^3$,  
$\epsilon^{0123}=-1$, 
${\rm Sp}\left (\gamma^5\gamma^{\mu}\gamma^{\nu}\gamma^{\alpha}\gamma^{\beta}\right )
=4i\epsilon^{\mu\nu\alpha\beta}$, $F_{\mu\nu}={\partial_\mu A_\nu-\partial_\nu A_\mu}$.
$\epsilon_{abcd}=\epsilon_{\alpha\beta\mu\nu}a^\alpha b^\beta c^\mu d^\nu$ for any
4-vectors $a,b,c,d$.}
\begin{eqnarray}
\partial^\nu (\bar q\gamma_\nu\gamma_5 q)=2im_q \bar q\gamma_5 q+N_c\frac{(eQ_q)^2}{16\pi^2} 
F\tilde F, \qquad \tilde F_{\mu\nu}=\epsilon_{\mu\nu\alpha\beta}F^{\alpha\beta}.  
\end{eqnarray}
with $F_{\mu\nu}$ the electromagnetic field tensor. This modification of the 
classical equation of motion accounts for the fact that 
the form factor $G^\gamma$ defined by the 2-photon matrix element 
\begin{eqnarray}
\label{Ggamma}
\langle \gamma(q_1)\gamma(q_2)|\partial^\nu (\bar q\gamma_\nu \gamma_5 q)|0\rangle
=e^2\epsilon_{q_1 \epsilon^*_1 q_2 \epsilon^*_2}G^\gamma(p^2,q_1^2,q_2^2),  
\end{eqnarray}
does not vanish for $m_q=0$ but turns out to be a constant independent of the current momentum $p=q_1+q_2$: 
\begin{eqnarray}
\label{abj-anomaly}
G^\gamma(p^2,q_1^2=q_2^2=0)=-{2N_c(Q_q)^2}/{4\pi^2}. 
\end{eqnarray} 
In this letter we study the properties of axial currents when one of 
the photons, $\gamma(q_2)$, is replaced by a vector meson $V(q_2)$, e. g. a $\rho$-meson.   
We demonstrate that the form factor $G^V$ defined according to the relation
\begin{eqnarray}
\label{GV}
\langle \gamma(q_1)V(q_2)|\partial^\nu (\bar q'\gamma_\nu \gamma_5 q)|0\rangle
=e\epsilon_{q_1 \epsilon^*_1 q_2 \epsilon^*_2}G^V(p^2,q_1^2,q_2^2)   
\end{eqnarray}
has also an anomalous behavior and does not vanish for massless quarks. 
This occurs in spite of the fact that the vector meson is a bound $q\bar q$ state and 
the corresponding loop graph has no ultraviolet divergence. 
Moreover, the anomalous behavior is observed for both, the neutral and the charged 
axial-vector currents. 

The classical equation of motion reads 
\begin{eqnarray}
\label{class}
\partial^\nu (\bar q'\gamma_\nu\gamma_5 q)=i(m_q+m'_q)\bar q'\gamma_5 q
+e(Q_{q'}-Q_{q})A^\nu \bar q'\gamma_\nu\gamma_5 q,  
\end{eqnarray}
where $A^\nu$ is the electromagnetic field. The $\langle \gamma V|...|0\rangle$ matrix 
element of the second term on the r.h.s. of  (\ref{class}) 
vanishes to order $e$. Therefore, to this order, the classical equation of motion 
(\ref{class}) predicts $G^V=0$ for $m_{q'}=m_q=0$. 
We find however to order $e$ and for large $p^2$ 
\begin{eqnarray}
G^V\sim M_V f_V/p^2,    
\end{eqnarray}
where $M_V$ and $f_V$ denote the mass and the decay constant of the vector meson. 
Because of the dependence on $p^2$, the newly found deviation from the classical 
equation of motion for bound states corresponds to a non-local anomaly. 

We are interested in the region $|p^2|\gg m_q^2, \Lambda_{QCD}^2$, in which 
case the quarks in the triangle diagram 
have high momenta and their propagation can be treated perturbatively. 
We discuss in parallel the $\gamma\gamma$ and $\gamma V$ final states 
in order to show how the anomaly emerges in both cases. 
As in Ref. \cite{dz}, we consider the spectral representation for the axial-vector 
current itself before forming the divergence. In distinction to \cite{dz}, where the 
spectral representation in $p^2$ was considered, we use the spectral representation in the 
variable $q_2^2$. This allows us to take bound state properties into account. 

\subsection{The absorptive part of the triangle amplitude}
The amplitude of the single-flavor axial current between the vacuum and the 
two-photon and the photon-vector meson states,   
respectively, can be written in the form 
\begin{eqnarray}
\epsilon^{*\beta}(q_2)\epsilon^{*\alpha}(q_1)T_{\nu\alpha\beta}(q_1,q_2). 
\end{eqnarray}
The absorptive part $t_{\nu\alpha\beta}$ of $T_{\nu\alpha\beta}$ is calculated 
by setting the two quarks attached to the external particle with the momentum $q_2$ on the mass shell, 
see Fig 1. 
\begin{center}
\begin{figure}[hb]
\vspace{-.4cm}
\mbox{\epsfig{file=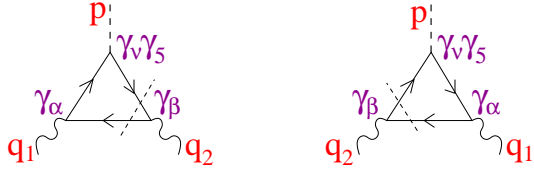,width=7cm}}
\vspace{.4cm}
\caption{Diagrams describing 
$\langle \gamma(q_1)\gamma(q_2)|\bar q \gamma_\nu\gamma_5 q|0\rangle$ 
and/or $\langle \gamma(q_1)V(q_2)|\bar q \gamma_\nu\gamma_5 q|0\rangle$, with 
$p=q_1+q_2$. 
The cut corresponds to the calculation of the absorptive part in the variable $q_2^2$.}
\end{figure}
\end{center}
$t_{\nu\alpha\beta}$ is our basis for the spectral representation of $T_{\nu\alpha\beta}$ in terms of the variable $q_2^2$. 
The coupling at the vertex $\beta$ is $\gamma_\beta e Q_q$ if the particle 2 is a photon, and 
$-\gamma_\beta g(q_2^2)/\sqrt{N_c}$ if it is a vector meson.\footnote{
The full vertex has the form \cite{m}
$\Gamma_\beta=-1/\sqrt{N_c}
[\gamma_\beta-\frac{1}{\sqrt{s}+2m}(k_1-k_2)_\beta/(\sqrt{s}+2m)]g((k_1+k_2)^2)$,  
but the term proportional to $(k_1-k_2)_\beta$ does not contribute to the trace. 
The overall $(-)$ sign is the standard choice of the phase of the vector meson wave
function which leads to a positive leptonic decay constant.}   
The coupling $g(q_2^2)$ will be further discussed below. By taking the trace and performing the 
integration over the internal momentum in the loop it is straighforward to obtain 
$t_{\nu\alpha\beta}$. The result is automatically gauge-invariant 
\begin{eqnarray}
q_1^\alpha t_{\nu\alpha\beta}(q_1,q_2)=0, \qquad q_2^\beta t_{\nu\alpha\beta}(q_1,q_2)=0. 
\end{eqnarray}
It is therefore possible to write the covariant decomposition of $t_{\nu\alpha\beta}(q_1,q_2)$ 
in terms of three invariant amplitudes 
\begin{eqnarray}
\label{imampl}
t_{\nu\alpha\beta}(q_1,q_2)=
-p_\nu \epsilon_{\alpha\beta q_1 q_2}ic_0
+(q_1^2\epsilon_{\nu\alpha\beta q_2}-q_{1\alpha}\epsilon_{\nu q_1\beta  q_2})ic_1 
+(q_2^2\epsilon_{\nu\beta\alpha q_1}-q_{2\beta} \epsilon_{\nu q_2\alpha q_1})ic_2.    
\end{eqnarray}
This Lorentz structure is chosen in such a way that no kinematical singularities appear. 
We take $\gamma(q_1)$ to be a real photon, $q_1^2=0$. Hence, the term 
containing the invariant amplitude 
$c_1$ does not contribute to the divergence of the current.  
Setting in addition $m_q=0$ one obtains for $c_0$ and $c_2$ with 
$s=q_2^2$
\begin{eqnarray}
\label{im-inv}
c_0(p^2,s)=-\frac{\zeta(s)}{4\pi}\frac{s}{(s-p^2)^2}, \qquad
c_2(p^2,s)=-\frac{\zeta(s)}{4\pi}\frac{p^2}{(s-p^2)^2},   
\end{eqnarray} 
where $\zeta(s)=2N_c Q_q^2 \theta (s)$ for the $\gamma\gamma$ process and 
$\zeta(s)=-2\sqrt{N_c} Q_q g(s)\theta (s)$ for the  $\gamma V$ process.  

Clearly, the absorptive part $t_{\nu\alpha\beta}(q_1,q_2)$ of the axial-vector current 
matrix element respects the classical equation of motion, that is
\begin{eqnarray}
\label{eq1}
p^2\, c_0(p^2,s)-s\,c_2(p^2,s)=0.  
\end{eqnarray}
 
\subsection{The triangle amplitude and its divergence}
The full amplitude $T_{\nu\alpha\beta}(q_1,q_2)$ has the same Lorentz structure as its absorptive 
part 
\begin{eqnarray}
\label{ampl}
T_{\nu\alpha\beta}(q_1,q_2)=
-p_\nu \epsilon_{\alpha\beta q_1 q_2}iC_0
+(q_1^2\epsilon_{\nu\alpha\beta q_2}-q_{1\alpha}\epsilon_{\nu q_1\beta  q_2})iC_1 
+(q_2^2\epsilon_{\nu\beta\alpha q_1}-q_{2\beta} \epsilon_{\nu q_2\alpha q_1})iC_2.    
\end{eqnarray}
The absence of any contact terms in $T_{\nu\alpha\beta}$ can be verified by 
reducing out one of the photons and using the conservation of the electromagnetic current. 
The invariant amplitudes $C_i$ can be represented by the following dispersion integrals 
\begin{eqnarray}
C_i(p^2,q_1^2=0,q_2^2)=\frac{1}{\pi}\int_{0}^{\infty}\frac{c_i(p^2,s)}{s-q_2^2-i0}ds. 
\end{eqnarray}
All the integrals converge and thus need no subtraction.

Taking the divergence of $T_{\nu\alpha\beta}$ we find 
\begin{eqnarray}
ip^\nu T_{\nu\alpha\beta}=
-\frac{1}{\pi}\left\{p^2 \int_{0}^{\infty}\frac{c_0(p^2,s)}{s-q_2^2}ds-
q_2^2 \int_{0}^{\infty}\frac{c_2(p^2,s)}{s-q_2^2}ds \right\}\epsilon_{q_1 \alpha q_2 \beta}. 
\end{eqnarray}
The form factor $G$ defined in Eqs. (\ref{Ggamma}) and (\ref{GV}) now reads  
\begin{eqnarray}
\label{Gint}
G(p^2,q_2^2)=\frac{p^2}{4\pi^2}\int_{0}^{\infty}\frac{\zeta(s)}{(s-p^2)^2}ds
\end{eqnarray}
In the case of the $\gamma\gamma$ process $\zeta(s)$ is a constant. The integral can be performed and 
gives the well-known value shown in Eq (\ref{abj-anomaly}). 
In the case of the $\gamma V$ matrix element the integrals converge even better since $g(s)$ which appears in $\zeta(s)$ descibes the spatial size of the vector meson. We conclude from 
Eq. (\ref{Gint}) that the divergence of the axial-vector current is nonzero for $m_q=0$ not only for the $\gamma \gamma $ but also for the 
$\gamma V$ final state! Namely, 
\begin{eqnarray}
\label{GVint}
G^V(p^2,q_2^2)=2\sqrt{N_c}e Q_q \frac{-p^2}{4\pi^2}\int_{0}^{\infty}\frac{g(s)}{(s-p^2)^2}ds. 
\end{eqnarray}
The behavior with respect to $p^2$ is however different from the $\gamma\gamma$ case and has the form 
$G^V(p^2)\sim 1/p^2$ for the large values of $p^2$ where our formula applies. 

For the transition to the $\gamma\rho$ (isospin-1) arising from the isovector axial current we obtain 
\begin{equation}
\label{GV1}
G^{\rho}=(Q_u+Q_d)\kappa \frac{f_\rho M_\rho}{p^2}, \qquad
\kappa=-\frac{\sqrt{N_c}}{4\pi^2}\frac{p^4}{f_\rho M_\rho}\int_{0}^{\infty}\frac{g(s)}{(s-p^2)^2}ds. 
\end{equation}
The parameter $\kappa$ in this equation is
non zero for $m_q=0$ and $|p^2|\to \infty$. $f_\rho$ is defined by the relation 
$\langle \rho^-|\bar d\gamma_\nu u|0\rangle=f_\rho M_\rho \epsilon^*_\nu$. 

Eq. (\ref{GV1}) takes into account the soft contribution to the form factor $G^{\rho}$.  
For large $|p^2|$ one should take care of the QCD evolution of the $\rho$-meson 
wave function 
from the soft scale $\mu^2\sim$ 1 GeV$^2$ to the scale $\mu^2 \sim |p^2|$.\footnote
{We want to  point here to the similarity of the form factor $G^\rho$ with the 
$\pi\gamma$ transition form factor $F_{\pi\gamma\gamma^*}(p^2)$. For a detailed analysis of 
the latter we refer to Ref. \cite{mr}. Likewise, the form factor $G^{\rho\rho}$ 
describing the amplitude $\langle \rho\rho|\partial^\nu (\bar q\gamma_\nu \gamma_5 q)|0\rangle$ has some common feature with 
the pion elastic form factor.}
This can be done most directly by expressing $\kappa$ in terms of the $\rho$-meson 
light-cone distribution amplitudes \cite{bb}   
\begin{eqnarray}
\nonumber
\langle\rho(q_2)|\bar d(x)\gamma_\lambda u(0)|0\rangle&=&
-iq_{2\lambda} (\epsilon^*x)f_\rho M_\rho\int\limits_0^1 du e^{iuq_2x}\Phi(u)
+\epsilon^*_\lambda f_\rho M_\rho \int\limits_0^1 du e^{iuq_2x}g_\perp^{(v)}(u),  
\\
\langle\rho(q_2)|\bar d(x)\gamma_\lambda \gamma_5u(0)|0\rangle&=&-\frac{1}{4}
\epsilon_{\lambda\eta\rho\sigma}\epsilon^{*\eta} q_2^\rho x^\sigma 
f_\rho M_\rho\int\limits_0^1 du e^{iuq_2x}g_\perp^{(a)}(u).   
\end{eqnarray} 
The diagrams of Fig 2 explain the procedure we follow: The quark propagator connects the 
axial current with the photon and the two distant space time points are bridged by the 
$\rho$ meson wave function. This leads to the following expression for $\kappa$  
\begin{center}
\begin{figure}[hb]
\vspace{-.6cm}
\mbox{\epsfig{file=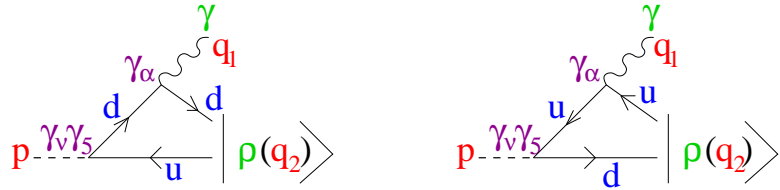,height=2.5cm}}
\vspace{.4cm}
\caption{The diagrams for the matrix element 
$\langle \gamma(q_1)\rho^-(q_2)|\bar d \gamma_\nu\gamma_5 u|0\rangle$, 
with two space time points bridged by the $\rho$ meson light-cone distribution amplitude.}
\end{figure}
\end{center}
\begin{eqnarray}
\kappa=-\int\limits_0^1 du \left[
 \frac{1+u}{4u^2}g_\perp^{(a)}
-\frac{1-u}{u}g_\perp^{(v)}
-\frac{1}{1-u}\Phi\right]. 
\end{eqnarray}
The leading-twist distribution amplitudes 
$g_\perp^{(a)}=6u(1-u)$, $g_\perp^{(v)}=\frac{3}{4}\left(1+(2u-1)^2\right)$, and 
$\Phi=\frac{3}{2}u(1-u)(2u-1)$ give the main contribution for large $p^2$ and lead 
to the value $\kappa=-3/2$. 
Corrections to this value are calculable 
in terms of the higher twist distribution amplitudes \cite{bb}. 

\vspace{.1cm}

Summing up, our results are as follows: 

\vspace{0.cm}
\noindent
1. The divergence of the axial vector current and thus the form factor $G$ does not 
vanish in the limit $m_q\to 0$. This holds for the $\gamma\rho$ final state as well 
as for the $\gamma\gamma$ final state. The observed effect for vector mesons requires 
a proper modification of the equation of motion for the axial current. 
For large momenta $p$ of the axial-vector current the corresponding 'bound state anomaly' 
can be described in terms of a non-local operator appearing at order $e$ 
(there are no local operators of the appropriate dimension): 
\begin{eqnarray}
\nonumber
&&{\rm Single\; flavor\;current}:
\\
&&\partial^\nu(\bar q\gamma_\nu\gamma_5 q)=2im\;\bar q \gamma_5 q +
\frac{(eQ_q)^2N_c}{16\pi^2}F\tilde F +
e\kappa\,Q_q\,\Box^{-1}
\left\{\partial_\mu(\bar q\gamma_\nu q)\cdot\tilde F^{\mu\nu}\right\}+O(\Box^{-2}). 
\\
{}\nonumber
\\
&&{\rm Isovector\; charged\; current}:
\nonumber\\
&&\partial_\nu(\bar d\gamma_\nu\gamma_5 u)=
i(m_u+m_d)\bar u \gamma_5 d 
-i e (Q_u-Q_d)\bar u \gamma_\nu\gamma_5 d\cdot A^\nu
+ e\kappa\frac{(Q_u+Q_d)}{2}\Box^{-1}\left\{
\partial_\mu(\bar u\gamma_\nu d)\cdot\tilde F_{\mu\nu}\right\}+O(\Box^{-2}). 
\end{eqnarray}
$\kappa$ can be expanded in a power series of $\alpha_s$.  
In leading order one obtains the value $\kappa=-3/2$. 

As pointed out in \cite{bmns}, the $\rho\gamma$ anomaly is important in 
rare decays: for instance, in $B\to\rho\gamma$ decays, it substantially 
corrects the weak annihilation amplitude which carries the CP violating 
phase.  

\vspace{0.cm}
\noindent
2. The amplitude for the $\rho\rho$ final state also stays finite for $m_q=0$. The
corresponding non-local anomalous term for the divergence of the isovector axial current 
$\partial_\nu(\bar q\gamma_\nu\gamma_5 {\tau^a}q)$ appears already at order $O(e^0)$. 
The operator structure of the anomalous term is more complicated in this case. 
One of the possible lowest-dimension operators which has a nonvanishing 
$\langle\rho\rho|...|0\rangle$ matrix element and thus will contribute to the 
anomalous term (in leading order in $1/p^2$) is the product of the two isovector 
tensor currents 
\begin{eqnarray}
\epsilon^{abc}\epsilon^{\mu\nu\alpha\beta}
\Box^{-1}\left(
\bar q \sigma_{\mu\nu}      {\tau^b}q\cdot  
\bar q \sigma_{\alpha\beta} {\tau^c}q\right). 
\end{eqnarray} 
Accordingly, for large $|p^2|$ the $\langle\rho\rho|...|0\rangle$ amplitude of the 
divergence of the axial-vector current is given by the factorized matrix element of the anomalous term and 
has a $1/p^2$ suppression 
\begin{eqnarray}
\langle \rho(q_1)\rho(q_2)|\partial^\nu(\bar d\gamma_\nu\gamma_5 u)|0\rangle=
\epsilon_{q_1 \epsilon^*_1 q_2 \epsilon^*_2}G^{\rho\rho},\qquad
G^{\rho\rho}\sim f_\rho^2 /p^2+O(1/p^4).  
\end{eqnarray} 

\vspace{0.cm}
\noindent 
3. We have illustrated the appearance of non-local anomaly due to 
vector mesons in QCD. This anomaly is of a general nature and should be present in 
any theory containing $J^P=1^-$ bound states. For example, the anomaly will also 
contribute to the generation of orthopositronium by the leptonic axial-vector current.
In contrast to the conventional local anomalies, there is no obvious cancellation 
of the non-local anomalies in the standard model. 

\vspace{.0cm}
\noindent{\it Acknowledgments}:  
{\small It is a pleasure to thank V. Braun, H. G. Dosch, O. Nachtmann, M. Neubert, 
O. P\`ene and V. I. Zakharov for discussions. D.M. was supported by the Alexander 
von Humboldt-Stiftung.}

\end{document}